\documentclass[prb,twocolumn,showpacs,preprintnumbers,amsmath,amssymb]{revtex4-1}

\usepackage{graphicx,rotating}
\usepackage{dcolumn}
\usepackage{bm}
\usepackage{epsfig}
\usepackage{longtable}

\begin{document}

\title {CVD Formation of Graphene on SiC Surface in Argon Atmosphere}

\author{Ma\l{}gorzata~Wierzbowska,$^1$, Adam Dominiak,$^2$ and Kamil Tokar$^1$} 

\affiliation{$^{1}$Institute of Theoretical Physics, Faculty of Physics,
University of Warsaw, ul. Ho\.za 69, 00-681 Warszawa, Poland,
$^{2}$Institute of Heat Engineering, Faculty of Power
and Aeronautical Engineering, Warsaw University of Technology,
ul. Nowowiejska 21/25, 00-665 Warszawa, Poland}

\begin{abstract}
We investigate the microscopic processes leading to graphene growth
by the chemical vapor deposition of propane in the argon atmosphere at the 
SiC surface. Experimentally, it is known that the presence of
argon fastens the dehydrogenation processes at the surface, 
in high temperature of about 2000~K. 
We perform ab-initio calculations, at zero temperature, to check whether
chemical reactions can explain this phenomenon. 
Density functional theory and supporting quantum chemistry methods 
qualitatively describe formation of the graphene wafers.
We find that the  4$H$-SiC(0001) surface exibits large catalytic effect
in the adsorption process of hydrocarbon molecules, this is also supported by
preliminary molecular dynamics results.
Existence of the ArH+ molecule, and an observation from the Raman spectra that 
the negative charge transfers into the SiC surface, would suggest that  
presence of argon atoms leads to a deprotonization on the surface,
which is necessary to obtain pure carbon add--layer.
But the zero-temperature description shows that the cold environment is 
insufficient to promote the argon-assisted surface cleaning.
\end{abstract}

\maketitle

\section{Introduction}

Recent progress in nanotechnology attracts much attention to 
graphene\cite{scientometric,guest,novel}.
Due to its elastic and electronic properties, this material is 
a very good candidate for novel devices with extraordinary 
features\cite{graphene-1,graphene-2,Karolcia,transistors,func}.
Preparation of pure, good quality, and large graphene wafers is of main
technological interest.
For many years, the SiC surfaces have been used for the graphene sheets 
growth in the epitaxy process by the Si sublimation\cite{graphene-formation}.
This method, however, introduces many defects and causes that graphene does  
not possess satisfactory electronic transport properties.
Structure of the epitaxial graphene and its interactions with the SiC
surface have been studied by Raman spectroscopy\cite{Wee}.

A new method of the epitaxy, by the chemical vapor deposition\cite{cvd1,cvd2} 
(CVD), is much less sensitive to the surface defects  
and enables to obtain high electron mobilities in the graphene layers  
(up to 1800 cm$^2$/Vs) and the grown wafers are large of even 150~mm in 
diameter\cite{Strupinski}.
Additionally, the graphene multilayers may be oriented in many 
stacking sequences\cite{sequence}.    
A difference between the graphene growth on SiC by the sublimation 
and the CVD process is pronounced\cite{jap}.
Very recent analysis of the experimental parameters in the CVD growth 
of the graphene and graphite sheets has been reported\cite{recent-cvd}.
The CVD method has been also applied on the silicon dioxide 
substrate (SiO$_2$)\cite{SiO2}, copper\cite{Cu,Cu2,Cu3}, nickel\cite{Ni} 
and iron\cite{Fe}. 
It enables to transfer graphene onto arbitrary substrates\cite{transfer}.

In the technology, the gas mixture of Ar and propane (C$_3$H$_8$), 
in a role of carbon precursor, is used as an ingredient
in the graphene epitaxy by the CVD process\cite{Strupinski,Ar1,Ar2}.
Propane is used in role of carbon precursor in graphene layer creation process.
It is desirable to understand how these compounds participate in 
the formation of the carbon layers, and especially, what is the mechanism 
of the removal of the hydrogen atoms from the Si-terminated 
SiC surface. The substrate surface  must be very clean in order to obtain 
a good quality graphene. A possible functionalisation of graphene with
the adsorbed hydrogen is a different issue\cite{graf-H,grafH}. 

In this work, the chemical reactions 
behind the CVD process are described, and mechanisms of the surface 
dehydrogenation are checked.
These mechanisms are closely related with the noble gases tendency to 
form the diatomic molecules with protons or, in specific conditions, 
with the neutral hydrogen atom.
The propane molecule, obviously, chemisorbs neither on the Si- nor 
the C-terminated 4$H$- or 6$H$-SiC surface 
(4$H$ and 6$H$ means the hexagonal crystal structure with the stacking period 
in the z-axis of 4 and 6, respectively).
This is because C$_3$H$_8$ is a molecule with all chemical bonds saturated. 
It will be shown that it is absolutely sufficient to remove 
whichever hydrogen atom from the propane molecule in order 
to adsorb such created specie at the SiC surface.
Further dehydrogenation of the molecule makes the adsorption stronger.
The problem that arises is what is the propane dehydrogenation 
initialization event, since there are only the saturated propane molecules 
present in the gas phase. 
Therefore, we start from investigations of the reactions 
with isolated propane in the gas phase, and later we model the following 
chemical reactions at the surface.
The possible role of argon in deprotonization reactions will be discussed. 
If the deprotonization scenario was true, then it would explain the Raman
measurements, whose show that the charge transfers from the adsorbates to
the SiC surface\cite{nasi}.

\section{Calculation details}

All calculations in this work were performed with the density functional theory 
(DFT)\cite{ks}, using the plane-wave package {\sc Quantum ESPRESSO}\cite{qe}. 
In order to verify the correctness of the results obtained by the DFT tool, 
used in the further studies, the solutions for the specific reactions were 
validated by the all-electron calculations with the quantum chemistry package
{\sc GAMESS}\cite{gamess}, which employs the localized basis sets and
treats the Coulomb interactions by means of the perturbative and/or 
the multiconfiguration methods. To get insight into mechanisms of 
hydrocarbon dehydrogenation on the surface, 
some preliminary molecular dynamics (MD) simulations at 
thermostat temperatures $\approx$ 1500~K were perfomed with 
the SIESTA code\cite{siesta} .  

\section{Results}

\subsection{Molecular reactions in the gas phase}  

Initially, we investigated a scenario with the
C$_3$H$_8$ $\longrightarrow$ C$_3$H$_7$ + H and
C$_3$H$_8$ $\longrightarrow$ C$_3$H$_6$ + H$_2$ reactions in vacuum.
The reaction energies presented in Table~\ref{molecules}
were obtained with the schemes: restricted
(open shell) Hartree-Fock, R(O)HF, without and with the second order
perturbation corrections for the dynamical correlations at
a level of the M\"oller-Plesset, MP2, method\cite{Szabo} (both by {\sc GAMESS})
and the DFT (by {\sc Quantum ESPRESSO}).
Additionally, the dissociation energies of H$_2$ were calculated to complete
a description of the reactions energetics.
Details of a set-up used in the calculations are given in
the supporting information.

\begin{table}
  \caption{
  Reaction energies (in eV), defined as the total energies of the products
 minus the total energies of the substrates, for the removal of hydrogen from
 propane.
 The parameter $r_e$ indicates the bond lengths (in \AA).
  C$_3$H$_7$ is obtained from C$_3$H$_8$ by a dissociation of H from the
middle C. And C$_3$H$_6$ is the propene molecule (hydrogens are dissociated
 from the middle and terminal C of propane).} 
  \label{molecules}
  \begin{tabular}{lcccc}
     \hline
      Reaction &  R(O)HF & MP2  & DFT  & exp.\cite{Szabo} \\
     \hline
       C$_3$H$_8$ $\longrightarrow$ C$_3$H$_7$$\;$ + H   &
       3.538 & 4.128 & 4.208 &  \-- \\
       C$_3$H$_7$ $\longrightarrow$ C$_3$H$_6$$\;$ + H   &
        1.671 &  1.341 & 1.850 & \-- \\
       C$_3$H$_8$ $\longrightarrow$ C$_3$H$_6$$\;$ + H$_2$   &
        1.662 & 1.450 & 1.563 & \-- \\
      H$_2$ $\longrightarrow$ H + H   &
         3.547 & 4.018 & 4.466 & 4.75 \\
       r$_e$ (H$_2$) &  0.730 & 0.738 & 0.753 & 0.741 \\
  ArH$^+$ $\longrightarrow$ Ar + H$^+$ & 2.825 & 3.048 & 4.151 & \-- \\
     r$_e$ (ArH$^+$) &  1.310 & 1.328 & 1.339 & \-- \\
    \hline
  \end{tabular}
\end{table}

Independently on the approximation level, the removal of one hydrogen from
propane needs a considerable amount of energy provided into 
the system (ca. 4 eV). 
In a case of the propene molecule (C$_3$H$_6$),  a part of the energetical 
cost has been consumed by a formation of the H$_2$ diatomic bond. 
Because of high Ar concentration in the gas mixture, 
it is quite plausible that argon atoms could assist in the above reactions 
leading to free the hydrogen atom or a proton.
This statement would be supported by the results of the quantum chemistry work 
on the dissociation of the HeH$^+$ molecule, led 
by Wolniewicz\cite{Wolniewicz}, where the separation of proton is 
an exothermic reaction with about 2.04 eV achieved. 
Thus, a possibility of argon binding with a proton in our system 
was calculated. The results are presented in Table~\ref{molecules}.
The energy gained from ArH$^+$ formation is smaller  
than the energy amount needed to remove one of the hydrogen atoms from 
the propane molecule. However, the hydrogen ionization energy is still
necessary to be taken into account. 
Some energy might be obtained from any of the kinetic processes,
which occur at high temperatures, or from the catalytic reaction with 
the SiC surface. Indeed, our preliminary results 
obtained with the MD support that fact. 
At averaged simulation temperature of Nos\'e thermostat around 1500 K,
C$_3$H$_8$ releases one hydrogen with the kinetic energy around 5 eV and
the remaining C$_3$H$_7$ moiety with the kinetic energy of 1.5 eV hits 
the surface zone and binds at the Si-site.  

Fragmentation of the propane molecules might be 
also caused by the electron transfer from the neutral propane
into the positively charged noble gases
(with unpaired electrons in the valence shell), as it has been
demonstrated experimentally\cite{futrell1,futrell2}.
On the other hand, at high temperatures in the range of 1300-1700~K, 
a similar decomposition of propane could be obtained without noble gases. 
This process was studied with IR laser absorption kinetic spectroscopy 
and discussed without any role of argon\cite{thermal}.
However, in the aforementioned experiment, the gas mixture
of C$_3$H$_8$ and Ar (as a major compound) has been used.    

\begin{figure*}
\includegraphics[scale=1.0]{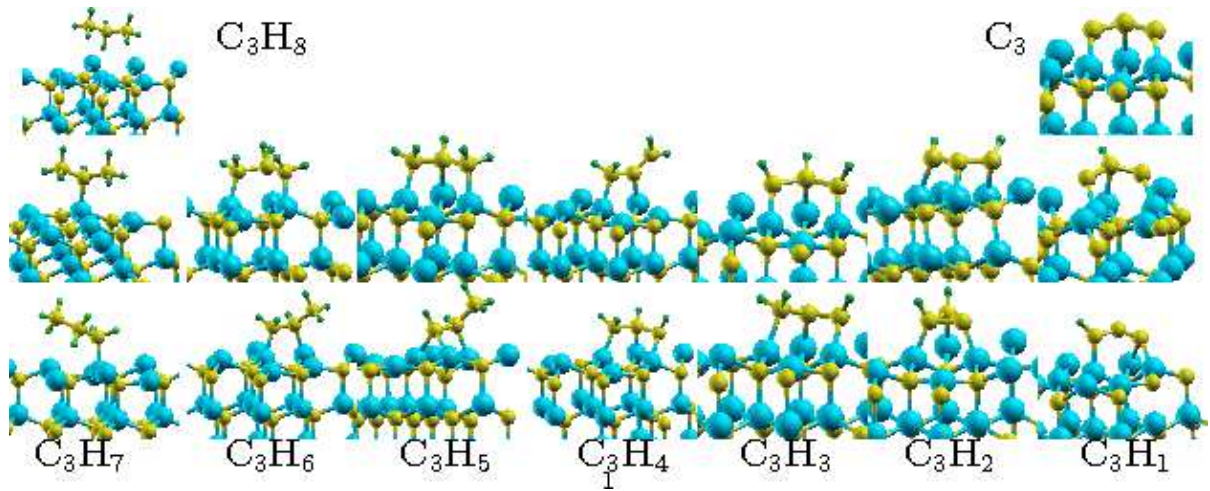}
\caption{ Adsorption geometries of propane and all transition
C$_3$H$_{8-n}$ species, where n=1,2,...,8  (up to the "naked" carbons)
at the Si-terminated SiC surface. The starting and final configurations,
C$_3$H$_8$ and C$_3$, are in the first row.
The second and third rows present the symmetric and the nonsymmetric cases,
respectively, for the descending number of hydrogen atoms from the left- to
the right-hand side.}
\label{fig:geoms}
\end{figure*}

To complete overview of the argon role in the investigated microscopic
mechanisms, it is needed to consider a possibility of the dehydrogenation
assisted by the formation of the neutral ArH molecule.  
Such process seems to be forbiden, since the noble gases have 
closed valence shell and are expected not to form molecules with other atoms.
We have checked, using the DFT and the ROHF methods, that indeed the neutral 
system ArH does not bind.
However, the Van der Waals complexes of Ar with propane
have been studied\cite{VdW-Ar}, and also the HeH$^+$ and ArH$^+$ 
charged molecules can be formed due to this type of interaction. 
Moreover, there are also known the diatomic molecules of NeH$^+$, KrH$^+$
and XeH$^+$ with the corresponding dissociation energies 2.08, 4.35
and 4.32 eV,\cite{Huber} respectively. 
Even more interesting are the molecules containing the noble gases and
some other atoms, where one or more ingredients are in the excited state.
It is known from the experiment that the molecule HArF occurs as
stable\cite{HArF}  and existence of HArCl and HHeF molecules have been 
predicted theoretically\cite{HArCl,HHeF} to be stable too.
Recently, the next two new molecules FArCCH and FArSiF$_3$ have also
been proposed\cite{ArSi}.

The crucial information for our investigations of argon role comes from
the multiconfigurational calculations for a dissociation of
the ArH$^*$ molecule in the excited state, performed by Vance and
Gallup\cite{Gallup}.
The main results of the work mentioned above are summarized in the
supporting information. Focusing on those data, we suppose that it is
impossible that argon could build a diatomic molecule
with neutral hydrogen in our system. This is because the curve minima, in
the dissociation channels of the excited argon, are shallow 
with 1-1.5 eV energy. 
This energy is much less than the hydrogen binding energy to 
the surface or hydrocarbon, and the argon excitations are
about 11.5 and 11.7 eV.
Such energy excitations of the system cannot be accessible 
on this scale without a strong laser beam.

\subsection{Adsorption at the surface}

\begin{figure}
\epsfxsize=9.5cm
\includegraphics[scale=0.25,angle=0.0]{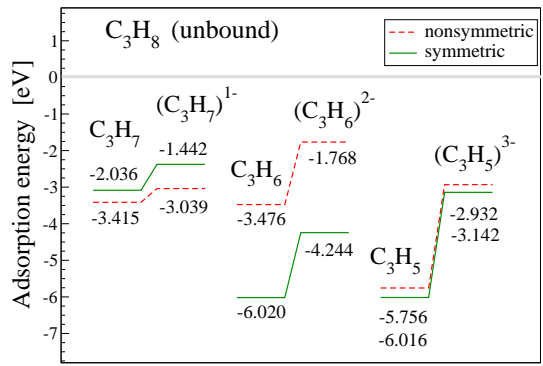}
\caption{Adsorption energies of the first three neutral and charged 
species at the surface obtained from a removal of the hydrogens from propane.}
\label{fig:ads}
\end{figure}

Assuming that, in a high temperature process, 
one hydrogen is removed from propane, the C$_3$H$_7$ system 
can be adsorbed at the surface. 
Two possibilities of creating such specie were defined: 
by 1) symmetric or 2) nonsymmetric removal of the hydrogen atom from 
the original hydrocarbon molecule. 
Since an adsorption at the 4$H$-SiC(0001) surface occurs for the both cases,
the symmetric CH$_3$-CH-CH$_3$ molecule and 
the nonsymmetric CH$_2$-CH$_2$-CH$_3$ molecule, further removals of  
the hydrogen atoms were considered and the adsorption energies 
were calculated.
Following this procedure, the adsorption of a series of the species 
C$_3$H$_{8-n}$, with n=1,...,7, was calculated. 
Finally, the hydrogen-free system, C$_3$, was adsorbed at the 4$H$-SiC(0001) 
surface. This type of hydrocarbon molecular residues might serve as precursors
for the graphene layer or a graphitic buffer layer\cite{buffer}. 
The studied adsorbent species build one, two or three valence bonds 
with the Si-terminated SiC surface. For any studied molecule, the
bond order formed with the surface atoms 
is strongly dependent on the specie--surface geometry and 
on number of hydrogens.
Some of the adsorbent created C-C bonds have a double bond character. 
The relaxed geometries of the adsorbed species are presented 
in Figure~\ref{fig:geoms}.

All calculated adsorption energies, except the C$_3$H$_8$ molecule, 
are negative, which means binding state. 
The modeled surface was considered to be metallic due to a saturation of
the surface with hydrogens\cite{Pollmann,Elwira}.

Adsorption energies were obtained from a formula valid for
the neutral and charged systems: 
\begin{equation}
 E_{\rm ads.} = E_{\rm slab+mol.} - E_{\rm slab} - E_{\rm mol.} - N\mu_{\rm e},
\end{equation}
where N is the number of additional electrons in the charged systems
(N$\neq$0 only in the cases presented in Figure~\ref{fig:ads}).
In the adsorption of charged molecules the total energies $E_{\rm slab+mol.}$
and $E_{\rm mol}$ were calculated with additional electrons, and 
the energy $E_{\rm slab}$ corresponds to the neutral surface.     
For the chemical potential of electrons, i.e. $\mu_{e}$,
we assumed the Fermi level of the pure slab (without the adsorbent) obtained
from the quadrature of the electronic density to the proper number of valence
electrons in the system with the used pseudopotentials. 
Modeling interactions in crystals, using the periodic supercells, 
introduces spurious interactions between periodic images
especially in the case of charged cells with the compensating charge 
uniform background. In order to take account of these effects,
we use the Makov and Payne method\cite{makov} 
implemented in the {\sc Quantum ESPRESSO} code. 
All geometries of the systems, taking a part in the adsorption process,
were optimized separatelly and non of the configurations was fixed.

The resulting values of the energies for the first three species: 
C$_3$H$_7$, C$_3$H$_6$ and C$_3$H$_5$  
are depicted in Figure~\ref{fig:ads}.  
Since it has been assumed, that the dehydrogenation could be assisted
by the ArH$^+$ molecule formation, the calculations for charged systems 
were also performed. It follows, that negatively charged species
bind weaker to the surface. The binding energy depends on the number 
of bonds, but also on the local surface strain induced by 
the adsorbed molecules. 
For example, the symmetric configuration of C$_3$H$_6$ group binds 
much stronger than the nonsymmetric one, due to a match of
the Si-terminated SiC surface lattice with the molecular C-C-C chain.
On the other hand, the C$_3$H$_7$ nonsymmetric 
molecule binds much stronger than the symmetric one,
because the CH$_3$ group in this specie is more distant from the surface 
when the terminal hydrogen is removed from propane.

 There exists a proposal of the charge transfer scenario
from the deprotonized site to the SiC surface states 
(which have extended delocalized character) 
assisted by formation of the ArH$^+$ molecule. 
The experimental data showed\cite{nasi}, that the charge distribution 
near the SiC surface is enhanced after the graphene layer adsorption.
Also the binding energy of ArH$^+$, of order 4.15 eV, is slightly
larger than the adsorption energy of the hydrogen atom at the Si-site of
the 4$H$-SiC(0001) surface, which amounts to 3.92 eV (from the DFT results).  
On the other hand, the energy of removal of a proton from the surface is
higher of the H ionization potential, about 13.6 eV, minus the work-function
of the SiC surface, circa 3.87 eV. Thus, the dissociation energy of 
a proton amounts to around 13.65 eV.  
This fact indicates that the zero temperature scenario with 
the argon-assisted surface chemical reactions does not take place. 

Further, the adsorption energies of the species with four or less hydrogens 
were compared with the adsorption energies of rich hydrogenated molecules.
In this comparison, the hydrogens dissociated from a molecule were adsorbed
at the surface Si-sites nearby the molecule
(somewhere in the middle of the primitive cell used in the calculations).
The adsorption sites were distant enough that 
the adsorbed species do not interact chemically. Although, in an indirect way
the surface deformations around the adsorbed molecules affect 
the adsorption energies. Thus, the final reaction was not just a
sum of two separate reactions with the surface. Such picture corresponds
to the experimental situation much better than a very separate adsorptions 
scenario, with hydrogens in the infinite distance from the molecule.  
The results of calculations for the aforementioned processes are included
in the supporting information, since the barriers were calculated via the
reactant in vaccum, and they do not include the catalytic role of the surface. 

\begin{table}
  \caption{Barriers (in eV) for the reactions below, which occur at the 
   SiC surface, for the symmetric and nonsymmetric adsorbates.
   The reaction directions are denoted by arrows ($\rightarrow$) and
   ($\leftarrow$) and defined by the differences between the
   highest energy configuration on the way from the left- to
   the right-hand side of given reaction and the energy of 
   the starting (for $\rightarrow$) or the final (for $\leftarrow$) 
   configuration, respectively, calculated within the NEB approach.} 
  \label{NEB}
  \begin{tabular}{lccccc}
     \hline
 Reaction  & $\;\;\;\;$ & 
            \multicolumn{2}{c}{symmetric} & 
            \multicolumn{2}{c}{nonsymmetric}  \\
         &  & $\rightarrow$ & $\leftarrow$ &
              $\rightarrow$ & $\leftarrow$  \\
     \hline
C$_3$H$_8$ $\longrightarrow$ C$_3$H$_7$ + H & & 
              0.15  &  6.91  &  0.70  &  3.40 \\
C$_3$H$_7$ $\longrightarrow$ C$_3$H$_6$ + H & &
              0.52  & 2.99   &  1.53 & 3.29  \\ 
C$_3$H$_6$ $\longrightarrow$ C$_3$H$_5$ + H & &
       1.06 &  2.40  &  1.45 &  2.47 \\
C$_3$H$_5$ $\longrightarrow$ C$_3$H$_4$ + H & &
       0.94 &  2.69  & 1.26 & 2.23  \\
C$_3$H$_4$ $\longrightarrow$ C$_3$H$_3$ + H & &
       0.79  & 1.49 & 0.98  & 2.14 \\
C$_3$H$_3$ $\longrightarrow$ C$_3$H$_2$ + H & &
       0.65 & 1.26 & 1.62 & 3.18 \\ 
C$_3$H$_2$ $\longrightarrow$ C$_3$H$_1$ + H & &
       0.43 & 1.04 & 1.34 &  3.02 \\
C$_3$H$_1$ $\longrightarrow$ C$_3$ + H & &       
      1.31 & 1.48 &  2.08 & 2.89 \\ 
    \hline
  \end{tabular}
\end{table}

\subsection{Energy barriers for the surface catalyzed dehydrogenations} 

Since the dehydrogenation processes which occur via the geometric 
configurations in vacuum show very high transition energies 
(see the supporting information), we calculated also
the minimum-energy paths for chosen reactions which take place at the surface.
In order to obtain the barriers for the reactions close to the surface, we
applied the climbing-image nudged-elastic-band method (NEB), 
implemented in the {\sc Quantum ESPRESSO} code\cite{qe}.
The results for chosen reactions are presented in Table~\ref{NEB}. 
Barrier energies are collected in columns corresponding to the symmetric 
and nonsymmetric geometries and to forward and backward reaction directions. 
The difference
between the highest energy on the reaction path and the energy of
the starting (or the final) geometric configuration gives the barrier for
the reaction forward $\rightarrow$ (or backward $\leftarrow$). The energy
differences between the starting and the final configurations can be obtained
from the differences ($\leftarrow$) - ($\rightarrow$). 
The barriers obtained on the minimum-energy path are not high. 
This implies, that the surface acts as a strong catalyzer in 
the dehydrogenation process of the hydrocarbon molecules.

The preliminary MD simulations of  processes after the adsorption of 
C$_3$H$_7$ show also cascade of dissociations. 
First, the released hydrogen from C$_3$H$_8$, or some other H from 
the atmosphere, collides with the remaining middle H of C$_3$H$_7$,
dissociating it and effectively creating H$_2$ outgoing back 
to the atmosphere. 
In the following dynamical evolution (time scale of 90-280 fs), one H atom from
the tail CH$_3$-group of remaining at the surface C$_3$H$_6$ specie 
is released, and immediately attracted to the surface Si-site 
neighbouring to the adsorption site of just deprotonized C$_3$H$_5$. \\

\section{Conclusions}

Role of argon and the SiC surface as catalysts in 
the dehydrogenation processes has been investigated. 
We started with a removal of one hydrogen atom from the C$_3$H$_8$ 
molecule and found it to be sufficient to initiate the adsorption reactions, 
which may continue with further dehydrogenation of molecules and more strong
binding, up to the C$_3$ moiety at the 4$H$-SiC(0001) surface. 
Barriers for the dehydrogenation of molecules at the surface,
with one of the reactants in vacuum and other at the surface, are
very high; except the first dehydrogenation of propane 
(see supporting information).
On the other hand, the barriers obtained on the minimum-energy paths 
for the hydrogen transfer from the adsorbed hydrocarbons onto 
the nearest Si-site at the SiC surface are rather low.
We conclude, that the SiC surface should act as a strong catalyzer 
in graphene epitaxy by the chemical vapor deposition process.

For the first time, we studied the chemical character of 
the dehydrogenation of molecules at the SiC slab, 
and not just the mechanical removing
of the H atoms by the floating gas. We check a microscopic mechanism for   
the dehydrogenation of the SiC surface, assisted by  
the binding reaction of a proton to argon forming the ArH$^+$ molecule. 
After this process, the electronic charge could remain on 
the surface\cite{nasi}. The zero-temperature description, however,
indicates that all proposed chemical reactions cannot occur without additional
processes caused by the high temperature kinetics or by a strong laser beam.  

Preliminary MD simulations without Ar in the atmosphere
above the surface, performed at high temperature of about 1500~K,
confirm the scenario with a cascade of dehydrogenations of the 
adsorbed hydrocarbons, and the fact that some of the dissociated hydrogens
remain at the surface. 

\section{Acknowledgement}

We would like to thank Jacek Majewski for many useful discussions.
This work has been supported by the European Funds for Regional Development
within the SICMAT Project (Contract No. UDA-POIG.01.03.01-14-155/09) and by
the European Union in the framework of European Social Fund through
{\em The Didactic Development Program of The Faculty of Power and Aeronautical
Engineering of The Warsaw University of Technology}.
Calculations have been performed in the Interdisciplinary Centre of
Mathematical and Computer Modeling (ICM) of the University of Warsaw
within the grant G47-7 and in Polish Infrastructure of
Informatic Support for Science in European Scientific Space (PL-Grid) 
within the projects nr POIG.02.03.00-00-028/08-00 and MRPO.01.02.00-12-479/02.

\end{document}